\documentclass[prd,preprint,superscriptaddress,showpacs,byrevtex]{revtex4}
\usepackage{bm}
\def\fsl#1{\setbox0=\hbox{$#1$}           
   \dimen0=\wd0                                 
   \setbox1=\hbox{/} \dimen1=\wd1               
   \ifdim\dimen0>\dimen1                        
      \rlap{\hbox to \dimen0{\hfil/\hfil}}      
      #1                                        
   \else                                        
      \rlap{\hbox to \dimen1{\hfil$#1$\hfil}}   
      /                                         
   \fi}                                         %
\newcommand{\be}{\begin{equation}}
\newcommand{\ee}{\end{equation}}
\newcommand{\bea}{\begin{eqnarray}}
\newcommand{\eea}{\end{eqnarray}}
\newcommand{\beq}{\begin{equation}}
\newcommand{\eeq}{\end{equation}}
\newcommand{\beqs}{\begin{eqnarray}}
\newcommand{\eeqs}{\end{eqnarray}}

\begin{document}
\title{Jet Quenching at RHIC and LHC and the Fragmentation Function in Vacuum }
\author{Gouranga C Nayak }\thanks{G. C. Nayak was affiliated with C. N. Yang Institute for Theoretical Physics in 2004-2007.}
\affiliation{ C. N. Yang Institute for Theoretical Physics, Stony Brook University, Stony Brook NY, 11794-3840 USA}
\date{\today}
\begin{abstract}
The value of the parton to hadron fragmentation function in QCD in vacuum (for example from proton-proton collisions at high energy colliders) is directly/indirectly used in the literature to study the jet quenching and the hadron production from quark-gluon plasma at RHIC and LHC. In this paper we show that this is not possible because, unlike the perturbative propagator in non-equilibrium QCD, the parton to hadron fragmentation function is a non-perturbative quantity in QCD and hence it is not possible to decompose the fragmentation function in non-equilibrium QCD into the vacuum part and the medium part.
\end{abstract}
\maketitle
\pagestyle{plain}
\pagenumbering{arabic}
\section{Introduction}

Just after $10^{-4}$ seconds of the big-bang our universe was filled with a state of matter known as the quark-gluon plasma (QGP). Hence it is important to recreate this early
universe scenario in the laboratory by producing quark-gluon plasma at the high energy heavy-ion colliders at RHIC and LHC. The hadronic matter undergoes phase transition to the  quark-gluon plasma phase at temperature $\ge$ 200 MeV ($10^{12}$ $^o$K).
The center of mass energy of two gold nuclei at RHIC is $\sqrt{s}$ = 200 GeV and the center of mass energy of two lead nuclei at LHC is $\sqrt{s}$ = 5.5 TeV. At such high energies the two nuclei at RHIC and LHC travel almost at speed of light.

Due to confinement in QCD the direct detection of quark-gluon plasma is not possible. The indirect signatures of quark-gluon plasma detection are: 1) $j/\psi$ suppression, 2)
dilepton production, 3) strangeness enhancement, 4) direct photon production and 5) jet quenching. In this paper we will investigate the jet quenching signature of the quark-gluon plasma detection at RHIC and LHC in detail.

In QCD in vacuum (for example in proton-proton collisions at high energy colliders) the back-to-back jets are experimentally observed. However, the jet quenching is observed at the high energy heavy-ion colliders at RHIC and LHC. The PHENIX \cite{phenix} and the STAR \cite{star} experiments at RHIC first reported the jet quenching in Au-Au collisions
at $\sqrt{s}$ = 200 GeV. Similarly the ATLAS \cite{atlas}, the CMS \cite{cms} and the ALICE \cite{alice} experiments at LHC have observed the jet quenching in Pb-Pb collisions at $\sqrt{s}$ = 2.76 TeV.

Since the jet quenching is experimentally observed at high energy heavy-ion colliders at RHIC and LHC, the main challenge for us is to theoretically understand these jet quenching from the first principle calculation by using QCD.

Since the two nuclei at RHIC and LHC travel almost at speed of light the longitudinal momenta of the partons inside the nuclei are much larger than their transverse momenta. These non-isotropy in momentum distribution creates the non-equilibrium quark-gluon plasma just after the nuclear collisions at RHIC and LHC. Many more secondary partonic collisions are necessary to bring the system into equilibrium. However, the hadronization time scale in QCD is very small ($\sim 10^{-24}$ seconds) which makes it unlikely that there are many more secondary partonic collisions to bring the system into equilibrium at RHIC and LHC. Hence the quark-gluon plasma at RHIC and LHC may be in non-equilibrium. This problem does not exist in QCD in vacuum, for example, in proton-proton collisions at high energy colliders.

In addition to this, there is another problem that needs attention which is the soft parton production at RHIC and LHC. This is because the soft partons play a major role in determining the bulk properties of quark-gluon plasma at RHIC and LHC. But the soft parton production can only be correctly calculated from first principle by using non-perturbative QCD which is not solved yet.

The third problem which is not solved yet is the hadronization from quark-gluon plasma using first principle calculation. This is because the hadron formation from parton involves non-perturbative QCD which is not solved yet.

These are the major difficult theoretical challenges we face today in order to do exact first principle calculation to study the production and detection of quark-gluon plasma at RHIC and LHC. Hence without doing exact first principle nonequilibrium-nonperturbative QCD calculation, the explanation of the experimental data at RHIC and LHC does not make sense. The first principle method to study nonequilibrium-nonperturbative QCD at RHIC and LHC is the closed-time path integral formalism in non-equilibrium QCD \cite{schw,keldysh,gr,co}.

Since some of the studies in the literature have explained experimental data at RHIC and LHC without doing exact first principle nonequilibrium-nonperturbative QCD calculation, let us briefly discuss the limitations of these studies.

First of all the lattice QCD at finite temperature \cite{lat} is used to study properties of quark-gluon plasma but this technique only works if the quark-gluon plasma is in equilibrium. Note that the definition of temperature does not exist in non-equilibrium. Hence the lattice QCD at finite temperature is not applicable for non-equilibrium quark-gluon plasma at RHIC and LHC.

Similarly the hydrodynamics is widely used to describe the experimental data at RHIC and LHC \cite{hyd} but the hydrodynamics can not correctly implement how partons form hadrons at RHIC and LHC which involves non-perturbative QCD which is not solved yet. As we will show in this paper the parton to hadron fragmentation function in QCD in vacuum (for example from proton-proton collisions) can not be directly/indirectly used to study hadron production from quark-gluon plasma at RHIC and LHC. In addition to this the hydrodynamics is not applicable in non-equilibrium. It should be mentioned here that even if hydrodynamics explains experimental data at RHIC and LHC it does not prove that the quark-gluon plasma at RHIC and LHC is in equilibrium. This is because, in order to prove that the quark-gluon plasma at RHIC and LHC is in equilibrium one has to prove that these experimental data can not be explained by non-equilibrium quark-gluon plasma.

AdS/CFT \cite{ads} and SYM plasma \cite{sym} have no connection with the real physics at RHIC and LHC because string theory and supersymmetry are not experimentally verified.

About the color glass condensate \cite{cgc} and the initial conditions at RHIC and LHC, as mentioned above, the soft parton production can only be correctly calculated from the first principle by using non-perturbative QCD which is not solved yet. The hard parton production can be calculated by using pQCD which is well known.

In the jet quenching study at RHIC and LHC the parton to hadron fragmentation function in QCD in vacuum (for example from proton-proton collisions at high energy colliders) is directly/indirectly used \cite{gw,gw1}. However, as we will show in this paper, this is not possible because the fragmentation function is a non-perturbative quantity in QCD and hence, unlike perturbative propagator in non-equilibrium QCD, it is not possible to decompose the fragmentation function in non-equilibrium QCD into the vacuum part and the medium part.

Hence for the reasons discussed above it is clear that there is no exact first principle theoretical calculation available at present to explain the experimental data at RHIC and LHC.

In proton-proton collisions at high energy colliders the parton to hadron fragmentation function in QCD in vacuum is used in the factorization formula to study hadron production \cite{css,n4,nkppn}. This factorization approach in QCD in vacuum needs to be extended to non-equilibrium QCD at RHIC and LHC to study hadrons production from non-equilibrium quark-gluon plasma.

Note that in non-equilibrium QCD the ground state is $|in>$ and in QCD in vacuum the ground state is $|0>$. The ground state $|in>$ in non-equilibrium QCD contains the information of both vacuum and medium, {\it i. e.}, when the distribution function $f({\vec p})=0$ then $|in>$ becomes equal to $|0>$ and we reproduce all the quantities in QCD in vacuum. For example, the perturbative propagator in non-equilibrium QCD is the sum of vacuum propagator and medium propagator, see eq. (\ref{gm}).

However, as we will show in this paper, since the parton to hadron fragmentation function is a non-perturbative quantity in QCD it is not possible to decompose the fragmentation function into the vacuum part and the medium part. Hence one can not use the fragmentation function in QCD in vacuum (for example from proton-proton collisions at high energy colliders) \cite{gw} in the medium modified DGLAP-like equation \cite{gw1} to study hadron production from quark-gluon plasma at RHIC and LHC. Any use of fragmentation function from QCD in vacuum (for example from proton-proton collisions at high energy colliders) will be inconsistent with the factorization theorem in non-equilibrium QCD (see sections V and VI). Hence we find that to be consistent with the factorization theorem in non-equilibrium QCD, the fragmentation function in non-equilibrium QCD should be used in the DGLAP-like equation in non-equilibrium QCD to study hadron production from quark-gluon plasma at RHIC and LHC.

This implies that the fragmentation function in QCD in vacuum (for example from proton-proton collisions at high energy colliders) can not be directlt/indirectly used to study the jet quenching and the hadron production from quark-gluon plasma at RHIC and LHC.

Note that when we say parton to hadron fragmentation function in non-equilibrium QCD one should keep in mind that the hadron $H$ is not in the non-equilibrium QCD medium. Only the parton is in the non-equilibrium QCD (in the in-state $|in>$) which fragments to hadrons. The hadron $H$ is formed in the out state $|H+X>$ where $X$ represents all other outgoing unobserved hadrons (see section VII).

The paper is organized as follows. In section II we describe QCD in non-equilibrium by using closed time path integral formalism. In section III we discuss the gauge invariant definition of the quark fragmentation function in non-equilibrium QCD. In section IV we discuss the gauge invariant definition of the gluon fragmentation function in non-equilibrium QCD. In section V we prove the factorization of infrared divergence of the quark fragmentation function in non-equilibrium QCD. In section VI we prove the factorization of infrared divergence of the gluon fragmentation function in non-equilibrium QCD. In section VII we show that the fragmentation function in QCD in vacuum (for example from proton-proton collisions at high energy colliders) can not be directly/indirectly used to study jet quenching and hadron production from quark-gluon plasma at RHIC and LHC. Section VIII contains conclusions.

\section{QCD in Non-Equilibrium Using Closed-Time Path Integral Formalism}

Quantum field theory in non-equilibrium can be studied by using closed-time path (CTP) formalism \cite{schw,keldysh}. The non-equilibrium QED is usually studied by using canonical quantization formalism. However, due to self interacting gluons and hadronization, the QCD in non-equilibrium can be best studied by using the closed-time path integral formalism.

We denote the (quantum) gluon field by $Q^{\mu d}(x)$ where $d=1,...,8$ is the color index. In the closed-time path integral formalism the generating functional in non-equilibrium QCD is given by \cite{gr,co}
\bea
&&Z[\xi_{+},{\bar \xi}_{+},\xi_-,{\bar \xi}_-,J_+,J_-,\rho]
=\int [d{\bar \psi}_{+}] [d{\bar \psi}_{-}] [d \psi_{+} ] [d\psi_{-}][dQ_+] [dQ_-]\nonumber \\
&&
\times {\rm det}(\frac{\delta \partial_\mu Q_+^{\mu d}}{\delta \omega_+^e})\times {\rm det}(\frac{\delta \partial_\mu Q_-^{\mu d}}{\delta \omega_-^e}) ~{\rm exp}[i\int d^4x [-\frac{1}{4}{F^d}_{\mu \nu}^2[Q_+]+\frac{1}{4}{F^d}_{\mu \nu}^2[Q_-] -\frac{1}{2 \alpha}(\partial_\mu Q_+^{\mu d })^2+\frac{1}{2 \alpha} (\partial_\mu Q_-^{\mu d })^2\nonumber \\
&&+{\bar \psi}_{+}  [i\gamma^\mu \partial_\mu -m +gT^d\gamma^\mu Q^d_{\mu +}]  \psi_{+}-{\bar \psi}_{-}  [i\gamma^\mu \partial_\mu -m +gT^d\gamma^\mu Q^d_{\mu -}]  \psi_{-} +J_+\cdot Q_+-J_-\cdot Q_-\nonumber \\
&&+{\bar \psi}_{+}\cdot \xi_{+} -{\bar \psi}_{-}\cdot \xi_{-}+{\bar \xi}_{+}\cdot \psi_{+} -{\bar \xi}_{-}\cdot \psi_{-}]] \times <\psi_+,{\bar \psi}_+,Q_+,t_{0}|~\rho~|t_{0},Q_-,{\bar \psi}_-,\psi_->
\label{zq}
\eea
where $+,-$ are the closed-time path indices and $\rho$ is the density of state in non-equilibrium QCD. In eq. (\ref{zq}) the state $|t_{0},Q_-,{\bar \psi}_-,\psi_->$ belong to time $t=t_{0}$ and
\bea
&&{F^d}_{\mu \nu}^2[Q]=[\partial_\mu Q_{\nu }^d-\partial_\nu Q_{\mu }^d+gf^{dcb} Q_{\mu }^cQ_{\nu }^b] \times [\partial^\mu Q^{\nu d}-\partial^\nu Q^{\mu d}+gf^{dae} Q^{\mu a}Q^{\nu e}] \nonumber \\
\label{f2}
\eea
where ${ \xi}_i$ and $J^{\mu d}$ are the external sources.

In the closed-time path integral formalism the generating functional in the background field method of non-equilibrium QCD is given by \cite{gr,co,th,ab,zu}
\bea
&&Z[A,\xi_{+},{\bar \xi}_{+},\xi_{-},{\bar \xi}_{-},J_+,J_-,\rho]
=\int [d{\bar \psi}_{+}] [d{\bar \psi}_{-}] [d \psi_{+} ] [d\psi_{-}][dQ_+] [dQ_-] \times {\rm det}(\frac{\delta G^d( Q_+)}{\delta \omega_+^e})\times {\rm det}(\frac{\delta G^d(Q_-)}{\delta \omega_-^e}) \nonumber \\
&&\times {\rm exp}[i\int d^4x [-\frac{1}{4}{F^d}_{\mu \nu}^2[Q_++A_+]+\frac{1}{4}{F^d}_{\mu \nu}^2[Q_-+A_-] -\frac{1}{2 \alpha}(G^d(Q_+))^2+\frac{1}{2 \alpha} (G^d(Q_-))^2\nonumber \\
&&+{\bar \psi}_{+}  [i\gamma^\mu \partial_\mu -m +gT^d\gamma^\mu (Q+A)^d_{\mu +}]  \psi_{+}-{\bar \psi}_{-}  [i\gamma^\mu \partial_\mu -m +gT^d\gamma^\mu (Q+A)^d_{\mu -}]  \psi_{-}+J_+\cdot Q_+-J_-\cdot Q_-\nonumber \\
&&+{\bar \psi}_{+}\cdot \xi_{+} -{\bar \psi}_{-}\cdot \xi_{-}+{\bar \xi}_{+}\cdot \psi_{+} -{\bar \xi}_{-}\cdot \psi_{-}]] \times <\psi_+,{\bar \psi}_+,Q_++A_+,t_{0}|~\rho~|t_{0},Q_-+A_-,{\bar \psi}_-,\psi_->
\label{zbq}
\eea
where
\bea
&& {F^d}_{\mu \nu}^2[Q+A]=[\partial_\mu [Q_{\nu }^d+A_{\nu }^d]-\partial_\nu [Q_{\mu }^d+A_{\mu }^d]+gf^{dcb} [Q_{ \mu }^c+A_{\mu }^c][Q_{\nu }^b+A_{\nu }^b]] \nonumber \\
&& \times [\partial^\mu [Q^{\nu d}+A^{\nu d}]-\partial^\nu [Q^{\mu d}+A^{\mu d }]+gf^{dae} [Q^{ \mu a}+A^{\mu a }][Q^{\nu e }+A^{\nu e}]].
\label{fb2}
\eea
In eq. (\ref{zbq}) the gauge fixing term
\bea
&& G^e(Q) =\partial_\mu Q^{\mu e} + gf^{ead} A_{\mu }^a Q^{\mu d}
\label{zgf}
\eea
depends on the background field $A^{\mu d}(x)$.

The type I gauge transformation is given by \cite{ab,zu}
\bea
&& T^e A'^{\mu e}(x) = \Phi(x) T^eA^{\mu e}(x)\Phi^{-1}(x) +\frac{1}{ig} [\partial^\mu \Phi(x)]\Phi^{-1}(x),~~~~~~~~~T^d Q'^{\mu d}(x) = \Phi(x) T^dQ^{\mu d}(x)\Phi^{-1}(x) \nonumber \\
\label{bgt}
\eea
where \cite{n4,nkppn,nkgf}
\bea
\Phi(x) =e^{igT^e \omega^e(x)}={\cal P}e^{-igT^d\int_0^\infty d\lambda l \cdot A^d(x+\lambda l)},~~~~~~~~~~l^2=0
\label{lw}
\eea
is the light-like gauge link (or the non-abelian phase) in the fundamental representation of SU(3).

\section{ Quark fragmentation function in non-equilibrium QCD }

As mentioned in the introduction the ground state $|in>$ at RHIC and LHC heavy-ion colliders is not the vacuum state $|0>$ any more due to the presence of the non-equilibrium QCD medium \cite{g1,g2,g3,g4,g5}. We denote the ground state in non-equilibrium QCD as $|in>$. Note that the non-equilibrium QCD ground state $|in>$ contains the information of both vacuum and medium. For example, when the medium is absent, {\it i. e.}, when the distribution function $f({\vec k})=0$ then the ground state $|in>$ becomes equal to the vacuum state $|0>$ and reproduces all the quantities in vacuum, see for example eq. (\ref{gm}).

For the quark with non-equilibrium (non-isotropic) distribution function $f_q({\vec k})$ we find that the gauge invariant definition of the quark to hadron fragmentation function which is consistent with the factorization of infrared divergences at all orders in coupling constant is given by
\bea
&& D_{H/q}(z,P_T)
= \frac{1}{12z~[1-f_q(k^+,k_T)]} \int dx^- \frac{d^{d-2}x_T}{(2\pi)^{d-1}}  e^{i{k}^+ x^- + i {P}_T \cdot x_T/z} \nonumber \\
&&{\rm Tr}_{\rm Dirac}~{\rm Tr}_{\rm color}[\gamma^+<in| \Phi^\dagger(x^-,x_T) {\psi}(x^-,x_T) a^\dagger_H(P^+,0_T)  a_H(P^+,0_T) \Phi(0){\bar \psi}(0) |in>]
\label{nqf}
\eea
where $\psi(x)$ is the quark field which fragments to hadron $H$ and $|in>$ is the ground state of the non-equilibrium QCD. In eq. (\ref{nqf}) the path ordered exponential
\bea
\Phi_{ij}(x)=\left[{\cal P}~ {\rm exp}[-ig\int_{0}^\infty d\lambda~ l \cdot A^d(x +l \lambda )~T^{d}]\right]_{ij}
\label{wf}
\eea
is the light-like gauge link in the fundamental representation of SU(3) where $l^\mu$ is the light-like four-velocity and $A^{\mu d}(x)$ is the SU(3) pure gauge background field.

\section{ Gluon fragmentation function in non-equilibrium QCD }

For the gluon with non-equilibrium (non-isotropic) distribution function $f_g({\vec k})$ we find that the gauge invariant definition of the gluon to hadron fragmentation function which is consistent with the factorization of infrared divergences at all orders in coupling constant is given by
\bea
&& D_{H/g}(z,P_T)
= \frac{k^+}{16z~[1+f_g(k^+,k_T)]} \int dx^- \frac{d^{d-2}x_T}{(2\pi)^{d-1}}  e^{i{k}^+ x^- + i {P}_T \cdot x_T/z} \nonumber \\
&&<in| Q^{\mu d}(x^-,x_T) \Phi^{(A)}(x^-,x_T)a^\dagger_H(P^+,0_T)  a_H(P^+,0_T)  \Phi^{(A)}(0)Q_\mu^d(0) |in>
\label{ngf}
\eea
where the quantum gluon field $Q^{\mu d}(x)$ fragments to hadron $H$. In eq. (\ref{ngf}) the path ordered exponential
\bea
\Phi^{(A)}_{ec}(x)=\left[{\cal P}~ e^{-ig\int_{0}^\infty \lambda~ l \cdot A^d(x +l \lambda )~T^{(A)d}}\right]_{ec},~~~~~~~~~~~T^{(A)d}_{ec}=-if^{dec}
\label{wa}
\eea
is the light-like gauge link in the adjoint representation of SU(3).

Note that the definitions of the quark and gluon to hadron fragmentation functions
in eqs. (\ref{nqf}) and (\ref{ngf}) are gauge invariant with respect to the gauge transformation \cite{th,zu,ab}
\bea
&&T^dA'^d_\mu(x) = U(x)T^dA^d_\mu(x) U^{-1}(x)+\frac{1}{ig}[\partial_\mu U(x)] U^{-1}(x),\nonumber \\
&&T^eQ'^e_\mu(x)=U(x)T^eQ^e_\mu(x) U^{-1}(x),~~~~~~~~~~~~~~~~U(x)=e^{igT^c\omega^c(x)}.
\label{gtf}
\eea
The special case $f_{q,g}(\vec{k})=\frac{1}{e^{\frac{k_0}{T}}\pm 1}$ corresponds to the quark-gluon plasma in equilibrium at the finite temperature $T$.
\section{Proof of factorization of infrared divergences of the Quark Fragmentation Function in non-equilibrium QCD}

In QCD the infrared (soft) divergence due to the real gluon emission occurs in the limit $k_0,k_1,k_2,k_3 \rightarrow 0$ where $k^\mu$ is the four-momentum of the emitted real gluon. The infrared (soft) divergence in QCD can be studied by using the eikonal approximation \cite{css,n4,nkppn,nkgf}
\bea
gT^d~\frac{q^\mu}{q \cdot k+i\epsilon }=gT^d~\frac{l^\mu}{l \cdot k+ i\epsilon }
\label{ek}
\eea
where $g$ is the coupling constant in QCD, $T^d$ is the generator in the gauge group SU(3) and $q^\mu$ is the four-momentum of the quark with four-velocity $l^\mu$. From eq. (\ref{ek}) we find that the eikonal contribution to the Feynman diagram of a real massless gluon emitted from a light-like quark ($l^2=0$) is given by
\bea
&& gT^d\int \frac{d^4k}{(2\pi)^4} \frac{l\cdot { A}^d(k)}{l\cdot k +i\epsilon } = igT^e\int_0^{\infty} d\lambda l\cdot {  A}^e(l\lambda)
\label{ek1}
\eea
where the Fourier transformation is given by
\bea
\int \frac{d^4k}{(2\pi)^4} { A}_\mu^{ e}(k) e^{ik \cdot x}={ A}_\mu^{ e}(x).
\label{ek2}
\eea
Similarly, when infinite number of real gluons are emitted from the light-like quark we find the corresponding eikonal contribution
\bea
{\cal P}~{\rm exp}[ig T^e\int_0^{\infty} d\lambda l\cdot { A}^e(l\lambda) ].
\label{po}
\eea
where ${\cal P}$ is the path ordering.

A light-like eikonal current produces pure gauge field at the spatial positions ${\vec l} \cdot {\vec x} \neq 0$ at the time $x_0=0$ in classical mechanics \cite{css,nj,ne} and in quantum field theory \cite{n4}.
For $ A^{\mu e}(\lambda l)$ in eq. (\ref{po}) we find that ${\vec l}\cdot {\vec x}=\lambda\neq 0$ which implies that the light-like quark finds the gluon field $A^{\mu e}(x)$ in eq. (\ref{po}) as the SU(3) pure gauge field \cite{n4,nkppn,nkgf}
\bea
T^eA^{\mu e} (x)= \frac{1}{ig}[\partial^\mu \Phi(x)] ~\Phi^{-1}(x),~~~~~~~~~~~~~\Phi(x)=e^{igT^d\omega^d(x)}={\cal P}~{\rm exp}[-ig T^e\int_0^{\infty} d\lambda l\cdot { A}^e(x+l\lambda) ]\nonumber \\
\label{pgf}
\eea
where $\Phi(x)$ is the light-like gauge link in the fundamental representation of SU(3) given by eq. (\ref{wf}).

Since the parton to hadron fragmentation function is a non-perturbative quantity in QCD, its property may not always be correctly studied by using perturbative QCD no matter how many orders of perturbation theory is used. The factorization of infrared (soft) divergences of the parton to hadron fragmentation function due to the soft gluons exchange with the light-like eikonal line at all orders in coupling constant in QCD can be studied by using the path integral formulation of the background field method of QCD in the presence of SU(3) pure gauge background field  \cite{n4,nkppn,nkgf}.

In the closed-time path integral formalism the generating functional in non-equilibrium QCD is given by eq. (\ref{zq}) and in the background field method of non-equilibrium QCD it is given by eq. (\ref{zbq}). From eq. (\ref{zq}) we find that the nonequilibrium-nonperturbative correlation function of the quark of type $<in|\psi^\dagger_r(x') \psi_s(x'')|in>$ in QCD is given by
\bea
&&<in|\psi^\dagger_r(x') \psi_s(x'')|in>=\int [dQ_+] [dQ_-][d{\bar \psi}_{+}] [d{\bar \psi}_{-}] [d \psi_{+} ] [d\psi_{-}] ~ \psi^\dagger_r(x') \psi_s(x'')\nonumber \\
&&\times {\rm det}(\frac{\delta \partial_\mu Q_+^{\mu d}}{\delta \omega_+^e})\times {\rm det}(\frac{\delta \partial_\mu Q_-^{\mu d}}{\delta \omega_-^e}) ~{\rm exp}[i\int d^4x [-\frac{1}{4}{F^d}_{\mu \nu}^2[Q_+]+\frac{1}{4}{F^d}_{\mu \nu}^2[Q_-] -\frac{1}{2 \alpha}(\partial_\mu Q_+^{\mu d })^2+\frac{1}{2 \alpha} (\partial_\mu Q_-^{\mu d})^2\nonumber \\
&&+{\bar \psi}_{+}  [i\gamma^\mu \partial_\mu -m +gT^d\gamma^\mu Q^d_{\mu +}]  \psi_{+}-{\bar \psi}_{-}  [i\gamma^\mu \partial_\mu -m +gT^d\gamma^\mu Q^d_{\mu -}]  \psi_{-}]] \nonumber \\
&&\times <Q_+,\psi_+,{\bar \psi}_+,t_{0}|~\rho~|t_{0},{\bar \psi}_-,\psi_-,Q_->
\label{cq}
\eea
where $r,s=+,-$ are the closed-time path indices.

Similarly from eq. (\ref{zbq}) we find that the non-equilibrium non-perturbative correlation function of the quark of type $<in|\psi^\dagger_r(x') \psi_s(x'')|in>$ in the background field method of QCD is given by
\bea
&&<in|\psi^\dagger_r(x')\psi_s(x'')|in>_A=\int [dQ_+] [dQ_-][d{\bar \psi}_{+}] [d{\bar \psi}_{-}] [d \psi_{+} ] [d\psi_{-}] \nonumber \\
&& \times \psi^\dagger_r(x') \psi_s(x'')\times {\rm det}(\frac{\delta G^d( Q_+)}{\delta \omega_+^e})\times {\rm det}(\frac{\delta G^d(Q_-)}{\delta \omega_-^e}) \nonumber \\
&&\times {\rm exp}[i\int d^4x [-\frac{1}{4}{F^d}_{\mu \nu}^2[Q_++A_+]+\frac{1}{4}{F^d}_{\mu \nu}^2[Q_-+A_-] -\frac{1}{2 \alpha}(G^d(Q_+))^2+\frac{1}{2 \alpha} (G^d(Q_-))^2\nonumber \\
&&+{\bar \psi}_{+}  [i\gamma^\mu \partial_\mu -m +gT^d\gamma^\mu (Q+A)^d_{\mu +}]  \psi_{+}-{\bar \psi}_{-}  [i\gamma^\mu \partial_\mu -m +gT^d\gamma^\mu (Q+A)^d_{\mu -}]  \psi_{-}]] \nonumber \\
&&\times <Q_++A_+,\psi_+,{\bar \psi}_+,t_{0}|~\rho~|t_{0},{\bar \psi}_-,\psi_-,Q_-+A_->.
\label{cbq0}
\eea
From eq. (\ref{cbq0}) we find
\bea
&&<in|\psi^\dagger_r(x') \Phi_r(x')\Phi^\dagger_s(x'')\psi_s(x'')|in>_A=\int [dQ_+] [dQ_-][d{\bar \psi}_{+}] [d{\bar \psi}_{-}] [d \psi_{+} ] [d\psi_{-}] \nonumber \\
&& \times \psi^\dagger_r(x')\Phi_r(x')\Phi^\dagger_s(x'') \psi_s(x'')\times {\rm det}(\frac{\delta G^d( Q_+)}{\delta \omega_+^e})\times {\rm det}(\frac{\delta G^d(Q_-)}{\delta \omega_-^e}) \nonumber \\
&&\times {\rm exp}[i\int d^4x [-\frac{1}{4}{F^d}_{\mu \nu}^2[Q_++A_+]+\frac{1}{4}{F^d}_{\mu \nu}^2[Q_-+A_-] -\frac{1}{2 \alpha}(G^d(Q_+))^2+\frac{1}{2 \alpha} (G^d(Q_-))^2\nonumber \\
&&+{\bar \psi}_{+}  [i\gamma^\mu \partial_\mu -m +gT^d\gamma^\mu (Q+A)^d_{\mu +}]  \psi_{+}-{\bar \psi}_{-}  [i\gamma^\mu \partial_\mu -m +gT^d\gamma^\mu (Q+A)^d_{\mu -}]  \psi_{-}]] \nonumber \\
&&\times <Q_++A_+,\psi_+,{\bar \psi}_+,t_{0}|~\rho~|t_{0},{\bar \psi}_-,\psi_-,Q_-+A_->
\label{cbq}
\eea
where $\Phi(x)$ is the gauge-link given by eq. (\ref{wf}).

We change the integration variable $Q \rightarrow Q-A$ inside the path integration in eq. (\ref{cbq}) to find
\bea
&&<in|\psi^\dagger_r(x') \Phi_r(x')\Phi^\dagger_s(x'')\psi_s(x'')|in>_A=\int [dQ_+] [dQ_-][d{\bar \psi}_{+}] [d{\bar \psi}_{-}] [d \psi_{+} ] [d\psi_{-}] \nonumber \\
&& \times \psi^\dagger_r(x')\Phi_r(x')\Phi^\dagger_s(x'') \psi_s(x'')\times {\rm det}(\frac{\delta G^d_f( Q_+)}{\delta \omega_+^e})\times {\rm det}(\frac{\delta G^d_f(Q_-)}{\delta \omega_-^e}) \nonumber \\
&&\times {\rm exp}[i\int d^4x [-\frac{1}{4}{F^d}_{\mu \nu}^2[Q_+]+\frac{1}{4}{F^d}_{\mu \nu}^2[Q_-] -\frac{1}{2 \alpha}(G^d_f(Q_+))^2+\frac{1}{2 \alpha} (G^d_f(Q_-))^2\nonumber \\
&&+{\bar \psi}_{+}  [i\gamma^\mu \partial_\mu -m +gT^d\gamma^\mu Q^d_{\mu +}]  \psi_{+}-{\bar \psi}_{-}  [i\gamma^\mu \partial_\mu -m +gT^d\gamma^\mu Q^d_{\mu -}]  \psi_{-}]] \nonumber \\
&&\times <Q_+,\psi_+,{\bar \psi}_+,t_{0}|~\rho~|t_{0},{\bar \psi}_-,\psi_-,Q_->
\label{cbq1}
\eea
where
\bea
&& G^e_f(Q_\pm) =\partial_\mu Q^{\mu e}_\pm + gf^{eda} A_{\mu \pm}^d Q^{\mu a}_\pm-\partial_\mu  A^{\mu e}_\pm \nonumber \\
&& T^e Q'^{\mu e}_\pm = \Phi_\pm T^eQ^{\mu e}_\pm \Phi^{-1}_\pm +\frac{1}{ig} (\partial^\mu \Phi_\pm)\Phi^{-1}_\pm.
\label{cbq2}
\eea
Since the change of integration variables do not change the value of the integration, the eq. (\ref{cbq1}) can be written in the form
\bea
&&<in|\psi^\dagger_r(x') \Phi_r(x')\Phi^\dagger_s(x'')\psi_s(x'')|in>_A=\int [dQ'_+] [dQ'_-][d{\bar \psi}'_{+}] [d{\bar \psi}'_{-}] [d \psi'_{+} ] [d\psi'_{-}] \nonumber \\
&& \times \psi'^\dagger_r(x')\Phi_r(x')\Phi^\dagger_s(x'') \psi'_s(x'')\times {\rm det}(\frac{\delta G^d_f( Q'_+)}{\delta \omega_+^e})\times {\rm det}(\frac{\delta G^d_f(Q'_-)}{\delta \omega_-^e}) \nonumber \\
&&\times {\rm exp}[i\int d^4x [-\frac{1}{4}{F^d}_{\mu \nu}^2[Q'_+]+\frac{1}{4}{F^d}_{\mu \nu}^2[Q'_-] -\frac{1}{2 \alpha}(G^d_f(Q'_+))^2+\frac{1}{2 \alpha} (G^d_f(Q'_-))^2\nonumber \\
&&+{\bar \psi}'_{+}  [i\gamma^\mu \partial_\mu -m +gT^d\gamma^\mu Q'^d_{\mu +}]  \psi'_{+}-{\bar \psi}'_{-}  [i\gamma^\mu \partial_\mu -m +gT^d\gamma^\mu Q'^d_{\mu -}]  \psi'_{-}]] \nonumber \\
&&\times <Q'_+,\psi'_+,{\bar \psi}'_+,t_{0}|~\rho~|t_{0},{\bar \psi}'_-,\psi'_-,Q_->.
\label{cbq3}
\eea

For the SU(3) pure gauge background field $A^{\mu e}(x)$ given by eq. (\ref{pgf}) we find from
\bea
\psi'_\pm(x) = \Phi_\pm(x) \psi_\pm(x)
\label{cbq4}
\eea
and from eq. (\ref{cbq2}) that \cite{n4,nkppn,nkgf}
\bea
&&[dQ'_r] =[dQ_r],~~~~~~~~~~~[d{\bar \psi}'_r] [d \psi'_r ]=[d{\bar \psi}_r] [d \psi_r ],~~~~~~~~~{F}^2[Q'_r]={F}^2[Q_r]\nonumber \\
&& (G_f^d(Q'_r))^2 = (\partial_\mu Q^{\mu d}_r(x))^2,~~~~~~~~~~~{\rm det} [\frac{\delta G_f^d(Q'_r)}{\delta \omega^e_r}] ={\rm det}[\frac{ \delta (\partial_\mu Q^{\mu d}_r(x))}{\delta \omega^e_r}] \nonumber \\
&&{\bar \psi}'_r [i\gamma^\mu \partial_\mu -m +gT^d\gamma^\mu Q'^d_{\mu r}] \psi'_r={\bar \psi}_r [i\gamma^\mu \partial_\mu -m +gT^d\gamma^\mu Q^d_{\mu r}]\psi_r.
\label{cbq5}
\eea
Using eqs. (\ref{cbq4}) and (\ref{cbq5}) in (\ref{cbq3}) we find
\bea
&&<in|\psi^\dagger_r(x') \Phi_r(x')\Phi^\dagger_s(x'')\psi_s(x'')|in>_A
=\int [dQ_+] [dQ_-][d{\bar \psi}_{+}] [d{\bar \psi}_{-}] [d \psi_{+} ] [d\psi_{-}] ~ \psi^\dagger_r(x') \psi_s(x'')\nonumber \\
&&\times {\rm det}(\frac{\delta \partial_\mu Q_+^{\mu d}}{\delta \omega_+^e})\times {\rm det}(\frac{\delta \partial_\mu Q_-^{\mu d}}{\delta \omega_-^e}) ~{\rm exp}[i\int d^4x [-\frac{1}{4}{F^d}_{\mu \nu}^2[Q_+]+\frac{1}{4}{F^d}_{\mu \nu}^2[Q_-] -\frac{1}{2 \alpha}(\partial_\mu Q_+^{\mu d })^2+\frac{1}{2 \alpha} (\partial_\mu Q_-^{\mu d})^2\nonumber \\
&&+{\bar \psi}_{+}  [i\gamma^\mu \partial_\mu -m +gT^d\gamma^\mu Q^d_{\mu +}]  \psi_{+}-{\bar \psi}_{-}  [i\gamma^\mu \partial_\mu -m +gT^d\gamma^\mu Q^d_{\mu -}]  \psi_{-}]] \nonumber \\
&&\times <Q_+,\psi_+,{\bar \psi}_+,t_{0}|~\rho~|t_{0},{\bar \psi}_-,\psi_-,Q_->
\label{cbq6}
\eea
Finally from eqs. (\ref{cbq6}) and (\ref{cq}) we find
\bea
<in|\psi^\dagger_r(x')\psi_s(x'')|in>=<in|\psi^\dagger_r(x') \Phi_r(x')\Phi^\dagger_s(x'')\psi_s(x'')|in>_A
\label{cqf}
\eea
which proves factorization of infrared divergences at all orders in coupling constant in non-equilibrium QCD where the light-like gauge link $\Phi(x)$ in the fundamental representation of SU(3) is given by eq. (\ref{wf}).

From eq. (\ref{cqf}) and \cite{nkfg} we find that the definition of the quark to hadron fragmentation function which is gauge invariant and is consistent with the factorization of infrared divergences at all orders in coupling constant in non-equilibrium QCD is given by eq. (\ref{nqf}).

\section{Proof of factorization of infrared divergences of the Gluon Fragmentation Function in non-equilibrium QCD}

Extending the above proof to the gluon case we find \cite{nkgf}
\bea
<in|Q^{\mu d}_r(x')Q^{\nu e}_s(x'')|in>=<in|Q^{\mu d}_r(x') \Phi^{(A)}_r(x')\Phi^{(A)}_s(x'')Q^{\nu e}_s(x'')|in>_A
\label{cgf}
\eea
which proves factorization of infrared divergences at all orders in coupling constant in non-equilibrium QCD where the light-like gauge link $\Phi^{(A)}(x)$ in the adjoint representation of SU(3) is given by eq. (\ref{wa}).

From eq. (\ref{cgf}) and \cite{nkgf} we find that the gluon to hadron fragmentation function which is gauge invariant and is consistent with the factorization of infrared divergences at all orders in coupling constant in non-equilibrium QCD is given by eq. (\ref{ngf}).

\section{ Jet quenching at RHIC and LHC and the fragmentation function in QCD in vacuum }

As mentioned earlier the parton to hadron fragmentation function is a non-perturbative quantity in QCD. Since the non-perturbative QCD is not solved yet, the value of the parton to hadron fragmentation function is extracted from the experiment. The parton to hadron fragmentation function is extracted from various fixed-target and collider experiments involving proton-proton ($pp$), electron-positron ($e^+e^-$) and lepton-hadron ($lh$) collisions etc.. Hence the value of the parton to hadron fragmentation function extracted from various experiments so far is in QCD in vacuum but not in non-equilibrium QCD. Since the parton to hadron fragmentation function in non-equilibrium QCD is not measured at any experiments, the parton to hadron fragmentation function in QCD in vacuum is directly/indirectly used \cite{gw}, for example, in the medium modified DGLAP equation \cite{gw1} to study hadron production from quark-gluon plasma at RHIC and LHC. However, as we will show in this section, this is not possible because in order to be consistent with the factorization theorem in non-equilibrium QCD one has to use the parton to hadron fragmentation function in non-equilibrium QCD in the DGLAP-like evolution equation in non-equilibrium QCD to study hadron production from quark-gluon plasma at RHIC and LHC.

As mentioned in the introduction, when we say the parton to hadron fragmentation function in non-equilibrium QCD one should keep in mind that the hadron $H$ is not in the non-equilibrium QCD medium. Only the parton is in the non-equilibrium QCD (in the in-state $|in>$) which fragments to hadrons. The hadron $H$ is formed in the out state
\bea
|H+X> = a^\dagger_H|X>
\eea
where $a^\dagger_H$ is the creation operator of the hadron $H$ and the state $|X>$ represents all other outgoing unobserved hadrons.

Using the factorization theorem in QCD in vacuum the inclusive hadron production cross section in proton-proton ($pp$) collisions at high energy colliders is given by the factorized formula in QCD in vacuum \cite{css,n4,nkppn}
\bea
\frac{d\sigma^H_{pp}}{dy dp_T^2}=\sum_j \int \frac{dz}{z^2} \frac{d{\hat \sigma}^{\rm V}_j}{dy dp_{Tj}^2} D^{\rm V}_{H/j}(z,Q)
\label{cv}
\eea
where the superscript symbol ${\rm V}$ stands for the vacuum. In eq. (\ref{cv}) the
$\frac{d{\hat \sigma}^{\rm V}_j}{dy dp_{Tj}^2}$ is the partonic level differential
cross section in QCD in vacuum in pp collisions and $D^{\rm V}_{H/j}(z,Q)$ is the parton to hadron fragmentation function in QCD in vacuum. The $Q^2$ evolution of the fragmentation function $D^{\rm V}_{H/j}(z,Q)$ in QCD in vacuum is given by the DGLAP evolution equation in vacuum \cite{dglap}
\bea
\frac{dD^{\rm V}_{H/j}(z,Q)}{d ln Q} = \frac{\alpha_s(Q)}{\pi} \sum_k \int_z^1 \frac{dz'}{z'} P^{\rm V}_{j\rightarrow kl}(z',Q) D^{\rm V}_{H/k}(\frac{z}{z'},Q)
\label{dg}
\eea
where $P^{\rm V}_{j\rightarrow kl}(z',Q)$ is the splitting function in QCD in vacuum with $j,k,l=q,{\bar q},g$.

Similar to QCD in vacuum in pp collisions at high energy colliders, we find that by
using the factorization theorem in non-equilibrium QCD [see eqs. (\ref{cqf}) and (\ref{cgf})] the inclusive hadron production cross section in heavy-ion ($AA$) collisions at RHIC and LHC is given by the factorized formula in non-equilibrium QCD
\bea
\frac{d\sigma^H_{AA}}{dy dp_T^2}=\sum_j \int \frac{dz}{z^2} \frac{d{\hat \sigma}_j}{dy dp_{Tj}^2} D_{H/j}(z,Q).
\label{cm}
\eea
In eq. (\ref{cm}) the $\frac{d{\hat \sigma}_j}{dy dp_{Tj}^2}$ is the partonic level differential cross section in non-equilibrium QCD in AA collisions at RHIC and LHC and $D_{H/j}(z,Q)$ is the parton to hadron fragmentation function in non-equilibrium QCD.
The $Q^2$ evolution of the fragmentation function $D_{H/j}(z,Q)$ in non-equilibrium QCD is
given by the DGLAP-like evolution equation in non-equilibrium QCD
\bea
\frac{dD_{H/j}(z,Q)}{d ln Q} = \frac{\alpha_s(Q)}{\pi} \sum_k \int_z^1 \frac{dz'}{z'} P_{j\rightarrow kl}(z',Q) D_{H/k}(\frac{z}{z'},Q)
\label{dgm}
\eea
where $P_{j\rightarrow kl}(z',Q)$ is the splitting function in non-equilibrium QCD \cite{nkd}.

In should be remembered that the factorization theorem is the crucial ingredient to derive DGLAP equation \cite{dglap}. Hence in order to be consistent with the factorization theorem in non-equilibrium QCD [see also eqs. (\ref{cqf}) and (\ref{cgf})] the
partonic level differential cross section $\frac{d{\hat \sigma}_j}{dy dp_{Tj}^2}$ in
eq. (\ref{cm}) is in non-equilibrium QCD and the fragmentation
function $D_{H/j}(z,Q)$ in eq. (\ref{cm}) is in non-equilibrium QCD which
are studied by using $|in>$ for the ground state in non-equilibrium QCD instead of the ground state $|0>$ in QCD in vacuum.

As mentioned earlier the ground state $|in>$ in non-equilibrium QCD contains both vacuum
and medium information, {\it i. e.}, when the distribution function $f({\vec p})=0$ then the ground state in non-equilibrium QCD $|in>$ becomes the vacuum state $|0>$ and
we reproduce all the quantities in the QCD in vacuum. For example, the leading order perturbative gluon propagator in non-equilibrium QCD is the sum of
the vacuum propagator and the medium propagator given by \cite{gr,co}
\bea
G^{\mu \nu}_{rs}(p) = [-ig^{\mu \nu} -i(\alpha -1) \frac{p^\mu p^\nu}{p^2}] ~G^{\rm vac}_{rs}(p) -iT^{\mu \nu}G^{\rm med}_{rs}(p)
\label{gm}
\eea
where the vacuum propagator $G^{\rm vac}_{rs}(p)$ is given by
\bea
G^{\rm vac}_{rs}(p)=
\left ( \begin{array}{cc}
\frac{1}{p^2+i\epsilon} & -2\pi \delta(p^2)\theta(-p_0) \\
-2\pi \delta(p^2)\theta(p_0) & -\frac{1}{p^2-i\epsilon}
\end{array} \right )
\label{vp}
\eea
and the medium propagator $G^{\rm med}_{rs}(p)$ is given by
\bea
G^{\rm med}_{rs}(p)= 2\pi \delta(p^2) f_g(\vec{p})
\left ( \begin{array}{cc}
1 & 1 \\
1 & 1
\end{array} \right ).
\label{mp}
\eea

However, since the parton to hadron fragmentation function is a non-perturbative quantity in QCD it is not possible to decompose the parton to hadron fragmentation function into the vacuum part and the medium part [see eqs. (\ref{nqf}) and (\ref{ngf})].
Hence we find that in order to be consistent with factorization theorem in non-equilibrium QCD, the fragmentation function in QCD in vacuum can not be used in the DGLAP-like equation in non-equilibrium QCD in eq. (\ref{dgm}) to study hadron production from quark-gluon plasma in AA collisions at RHIC and LHC. The parton to hadron fragmentation function in non-equilibrium QCD should be used in DGLAP-like equation in non-equilibrium QCD
in eq. (\ref{dgm}) to study hadron production from quark-gluon plasma in AA collisions at RHIC and LHC by using factorized formula as given by eq. (\ref{cm}) where $\frac{d{\hat \sigma}_i}{dy dp_{Tj}^2}$ is the partonic level differential cross section in non-equilibrium QCD and $D_{H/j}(z,Q)$ is the fragmentation function in non-equilibrium QCD.

Hence we find that the parton to hadron fragmentation function in QCD in vacuum (for example from proton-proton collisions at high energy colliders) can not be directly/indirectly used to study jet quenching and hadron production from quark-gluon plasma at high energy heavy-ion colliders at RHIC and LHC.

\section{Conclusions}
The value of the parton to hadron fragmentation function in QCD in vacuum (for example from proton-proton collisions at high energy colliders) is directly/indirectly used in the literature to study the jet quenching and the hadron production from quark-gluon plasma at RHIC and LHC. In this paper we have shown that this is not possible because, unlike the perturbative propagator in non-equilibrium QCD, the parton to hadron fragmentation function is a non-perturbative quantity in QCD and hence it is not possible to decompose the fragmentation function in non-equilibrium QCD into the vacuum part and the medium part.

Hence one finds that the parton to hadron fragmentation function in QCD in vacuum (for example from proton-proton collisions at high energy colliders) can not be directly/indirectly used to study jet quenching and hadron production from quark-gluon plasma at high energy heavy-ion colliders at RHIC and LHC.

\end{document}